\documentclass[letterpaper,twocolumn,10pt]{article}
\usepackage{usenix,epsfig,cite}
\usepackage{subfig}
\usepackage{graphicx}
\graphicspath{{.}}

\usepackage{url}

\newcommand{\myfig}[5]
{
\begin{figure}[t]
\begin{center}
\ifpdf
\includegraphics[width=#4\linewidth]{#1}
\else
\includegraphics[width=#4\linewidth]{#1}
\fi
\end{center}
\vspace{#5}
\caption{#2}\label{#3}
\vspace{-6pt}
\end{figure}
}

\newcommand{\mysubfloat}[5]
{
\subfloat[#2]{\label{#3}\includegraphics[width=#4\linewidth]{#1}\vspace{#5}}
}

\begin{document}

\date{}

\title{\Large \bf Glimmers: Resolving the Privacy/Trust Quagmire}

\author{
{\rm David Lie}\\
University of Toronto/Google
\and
{\rm Petros Maniatis}\\
Google
} %

\maketitle

\thispagestyle{empty}

\subsection*{Abstract}

Many successful services rely on trustworthy contributions from users.  To
establish that trust, such services often require access to privacy-sensitive
information from users, thus creating a conflict between privacy and trust.
Although it is likely impractical to expect both absolute privacy and 
trustworthiness at the same time, we argue that the current state of things,
where individual privacy is usually sacrificed at the altar of trustworthy services, can be
improved with a pragmatic \emph{Glimmer of Trust}, which allows services to validate user contributions in a trustworthy way without forfeiting user privacy.
We describe how trustworthy hardware such as Intel's SGX can be used client-side
-- in contrast to much recent work exploring SGX in cloud services -- to realize
the Glimmer architecture, and demonstrate how this realization is able to resolve the tension between privacy and trust in a variety of cases.

\begin{figure*}[t]
  \centering
  \mysubfloat{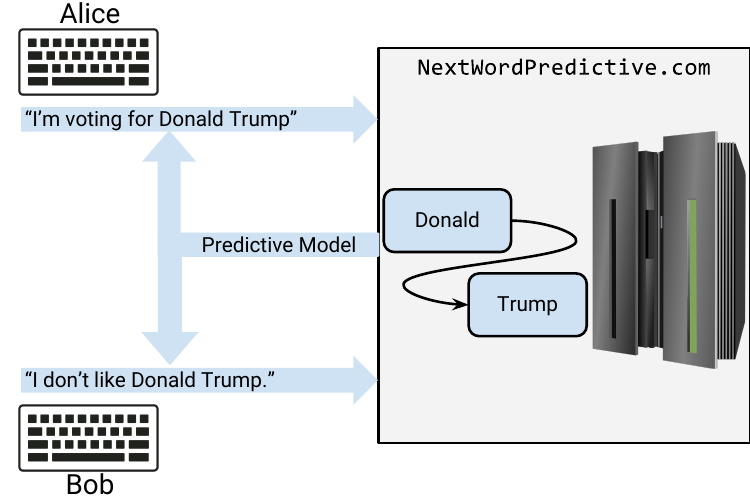}{A next-word predictive service for users' keyboards.}{fig:NextWordPredictive}{0.485}{-20pt}\quad
  \mysubfloat{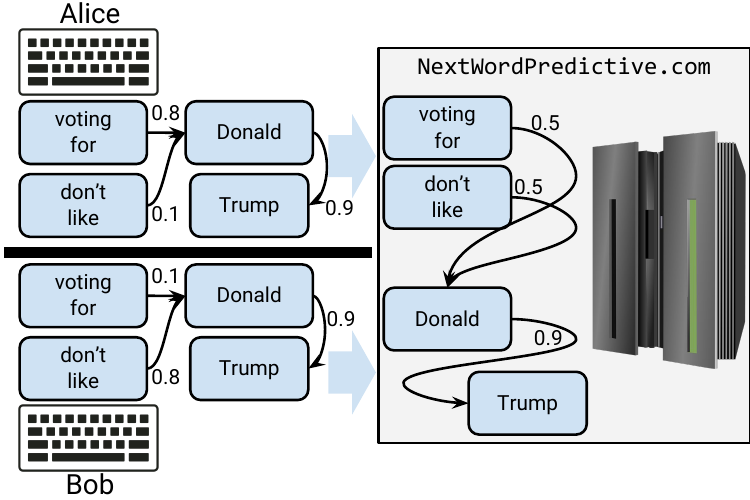}{Federated machine learning.}{fig:FML}{0.485}{-20pt}\quad
  \mysubfloat{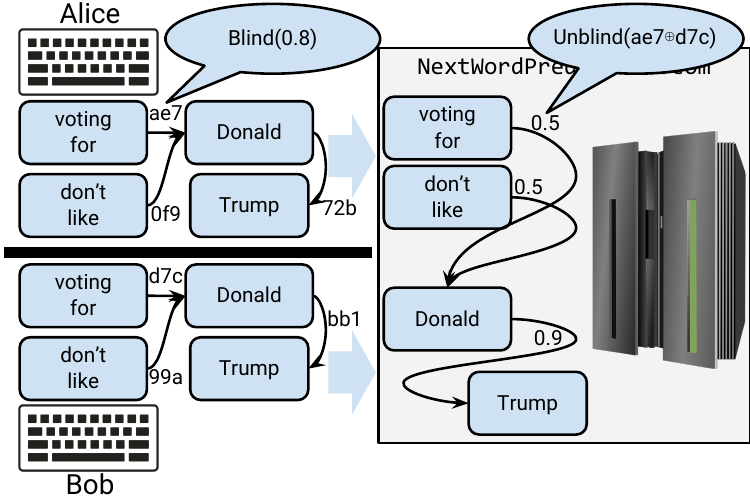}{Secure model aggregation.}{fig:SecAgg}{0.485}{-20pt}\quad
  \mysubfloat{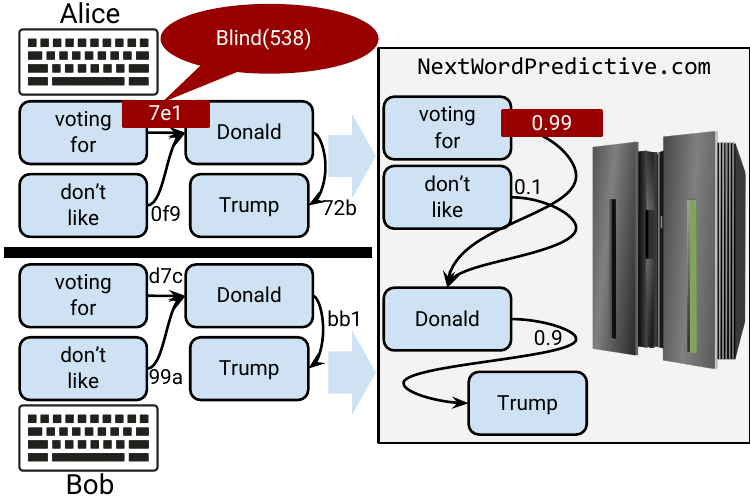}{Attack against secure model aggregation.}{fig:SecAggAttack}{0.485}{-20pt}
  \vspace{-6pt}
  \caption{A predictive keyboard service}
  \vspace{-10pt}
\end{figure*}

\section{Privacy and Cloud Services At Odds}\label{sec:intro}

More than ever, modern applications and services perform computation in a
distributed fashion.  Driven on one hand by the broad availability and adoption
of cloud services, and on the other by the shift to lightweight and
battery-constrained devices as clients, applications usually feature a
rich client backed by sophisticated computation and data storage at a service
provider.  Aside from the benefits of performance and energy efficiency,
cloud-backed applications also allow controlled sharing of information across
clients that leads to improved services, especially in the era of big-data
analytics and machine learning.  For example, a predictive keyboard on a client
device could benefit from a trained machine-learning model that uses inputs from
different users' keyboards (Figure~\ref{fig:NextWordPredictive}).  As current
topics (such as ``the world series'' or ``Donald Trump'') trend up -- because
many users type them on their keyboards in a short time-span -- an up-to-date
model can suggest ``Trump'' as the next word when Alice types ``Donald'', even
if she has never typed that name herself before.  Similar benefits exist for a
wide range of applications, such as image recognition, recommendation systems and activity detection.

Sadly, sharing of information, especially deeply personal information such as
key clicks, naturally introduces a tussle between information utility and user
privacy. If Alice's keyboard streams all her key clicks to a shared, predictive keyboard service, it will almost certainly reveal sensitive information
about her political beliefs. This is especially troubling, given that most
service providers release scant, if any, information about what
precisely they do with information they collect. For example, Bob's 
disparaging remarks enable a malicious service provider to
discriminate against him or aid in his prosecution. Even if the service provider
is well-meaning, compromising information can be subpoenaed by the
government to persecute Bob, stolen by hackers to extort him or damage the
service brand, or used by regulatory bodies to slam the service provider with
privacy-violation fines. Therefore, unmediated information sharing is
undesirable to users and services alike.

One way to alleviate this tussle is to process user data at the client, so that
it is never disclosed to the service in the
raw~\cite{shokri_privacy-preserving_2015, fml}. With Federated Machine
Learning~\cite{fml}, for example, every client computes a local partial model,
and the service sums those models together to generate a global one.  For
example, a simplistic keyboard model in Figure~\ref{fig:FML} associates a weight
between $0$ and $1$ for an ordered pair of words, without revealing the actual
sentences typed. However, learned models, even ones much more sophisticated than
our strawman illustration, can still reveal information about the raw inputs
used to train those models (e.g., machine-learning models can be
inverted~\cite{Fredrikson:2015:MIA:2810103.2813677}). To prevent partial models
from being inverted, users could contribute information using cryptographic
blinding~\cite{bonawitz2016practical} (Figure~\ref{fig:SecAgg}), or similar
techniques that enable accurate aggregation of a group of inputs while still
hiding the individual inputs.

Unfortunately, since these blinding or cryptographic techniques hide the inputs from the service, they afford malicious
users the ability to contribute forged or malicious inputs undetected, derailing the
quality of the shared service; even if specific validation checks existed (e.g.,
range checks, randomness checks, etc.), a client would have to be entrusted with
performing those checks before hiding the true inputs. For example, Alice could
contribute a blinded local model for her own keyboard sequences that has been
maliciously manipulated to over-weight her personal political convictions (i.e.,
contributing an illegal value of $538$ for one model parameter, violating the
valid rage of $[0,1])$, making it seem extremely popular beyond what a single
user might make it (Figure~\ref{fig:SecAggAttack}). When the service aggregates
the blinded local models together, it cannot detect such induced bias (because
of the blinding), and ends up with a catastrophically skewed global predictive
model towards inaccurate predictions, degrading the experience of all users. The
only way to range-check Alice's contribution is before blinding occurs, but the
service cannot trust Alice to do that faithfully. Even if the actual user
contributions are not themselves private, e.g., users photos associated with a
location on a mapping service, validating those contributions might require
access by service code to otherwise private data (e.g., location tracking
through GPS and ambient WiFi, to validate that the user did go to a claimed
location) at the device of the very user whose malice the validation is meant to
protect from.

Services affected by this trust-privacy trade-off have similar characteristics:
a) they consume user-contributed data to build global state benefiting all
users, b) service quality is highly dependent on the trustworthiness of data contributed by users, and c) they can only verify the legitimacy of user contributions through
direct access to sensitive user data (the contributions themselves, or 
contextual user information such as logs and other user activity).  Many of the previously mentioned services have these properties. For example, image recognition can benefit from local partial models trained on the private photos of users, but to verify that contributed photos are legitimate requires direct access to them before they are blinded; recommender services learn similarities among products from individual users'
registered likes, dislikes, and shopping habits, but detecting spurious reviews
requires access to individual users' purchasing history; activity-recognition
models improve from analyzing silhouettes and image structure from in-home
cameras, but checking that silhouettes are legitimate requires analysis of full
video streams captured at people's homes.  Ultimately, it seems then
that trust and privacy in such services is a zero-sum game -- an increase in
individual privacy results directly in a decrease in the amount of trust all the
users can collectively place in the service.

\section{A Glimmer of Trust In Between}\label{sec:problem}

\myfig{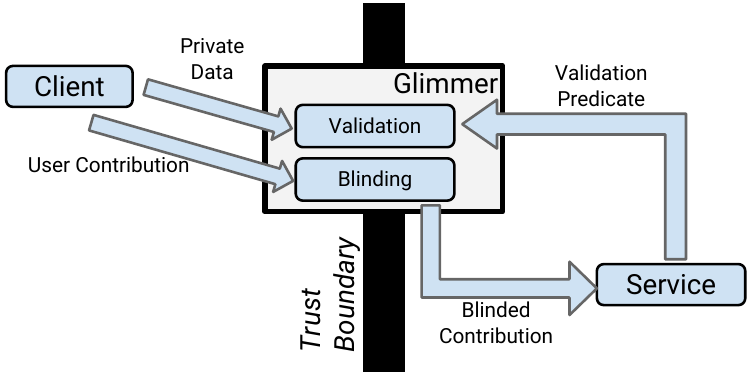}{Glimmer Architecture.}{fig:GlimmerArchitecture}{1.0}{-20pt}

To visualize this conundrum between privacy and trust, we introduce the concept
of a \emph{Trust Boundary} between the client and the service; no private data
should cross from the client to the service, but the correctness of client
contributions must be checked according to criteria set by the service on the
client's side.  The server cannot establish trust without access to the private
data, but at the same time the client does not want the server to access its
private data unrestricted.

Such ``air-gap'' problems can be solved with the introduction of a trusted
third party that performs {\em validation} on private client data before
submitting user contributions to the service.
We call our logical trusted third party a \emph{Glimmer of Trust}, or Glimmer for short,
since it performs very limited but essential trusted functionality: validation
of private data as specified by the service, followed by submission to the service
(Figure~\ref{fig:GlimmerArchitecture}).

We use the term validation loosely here to capture any validity predicate
entrusted upon the trusted third party; different validation predicates may
trade-off computational complexity for result accuracy.  For example, in the
next-word prediction service, range-checking model parameters ensures that Alice
cannot send a user contribution of $538$ when a value between $0$ and $1$ is
expected; however, she can still send arbitrary fictitious values within that
range that may not correspond to her actual keyboard activity. A more
sophisticated validator might instead observe actual keyboard behavior (a la
NAB~\cite{Gummadi:2009:NIS:1558977.1558998}) to match keyboard events to
reported model weights; or even observe CPU branches~\cite{vasudevan2011xtrec}
to identify a plausible execution of the model-construction code that produced
contributed partial results, as has been suggested for on-line game cheat
detection~\cite{Bethea:2008:SVC:2043628.2043633}.  While more invasive
validation increases the complexity and resources required by the Glimmer, it
also increases the adversary's cost to cheat undetected, since she now has to fabricate
keyboard activity or program executions that corroborate her deceptive inputs to
the service.

Regardless of the actual validation semantics, the Glimmer must satisfy certain
properties to be helpful. First, it must guarantee that it discards raw inputs
after processing, and that its outputs leak a bounded amount of information about private data, via encryption or aggregation ({\bf Input Confidentiality}). Second, it only endorses for use by the service those contributions that it has validated ({\bf Input Integrity}).

Having an actual third party performing the role of the Glimmer is, arguably,
the realization of this architecture. For example, the Electronic Frontier
Foundation (EFF), or a consortium of privacy advocacy organizations could, in
ensemble, perform validation and blinding, perhaps using multi-party
computation, or simpler threshold cryptography on inputs. However, the
deployment cost for such a solution would be high. In this vision paper, we
focus on using trustworthy hardware, because of its current broad availability,
as an implementation platform for Glimmers performing privacy-preserving
user-data validation.

\section{Glimmers on  SGX}

There are several instances of trustworthy hardware commonly available on
computing clients today, such as Intel TXT, AMD
SVM, ARM TrustZone, and Intel SGX.  In general, all these platforms provide a
hardware-enforced trusted execution environment (TEE), which can 
execute functionality isolated from any vulnerabilities or malicious code.

In this position paper we focus on realizing Glimmers using
Intel's SGX~\cite{sgx}.  SGX has spawned renewed interest in trusted computing,
with a number of server-side
uses~\cite{hunt_ryoan:_2016,schuster_vc3:_2015,arnautov_scone:_2016}.  We are
instead studying how SGX can be used on clients to realize Glimmers of
Trust~\footnote{Note that SGX is only available on client-class CPUs.}.  SGX provides
a TEE called an {\em enclave}. In addition to isolation, an SGX enclave also
supports {\em remote attestation}, which allows it to prove cryptographically to
a remote party that it is running correctly in a legitimate enclave, and {\em
sealed storage}, which allows it to encrypt data so that only it or other designated binaries, running in a legitimate enclave, can decrypt it.  Although very powerful, SGX
enclaves operate using limited resources, have no direct access to privileged
CPU operations such as IO, and must mediate system services via the untrusted
host OS. As a result, Glimmers implemented as enclaves must be simple and run mostly in isolation.

\myfig{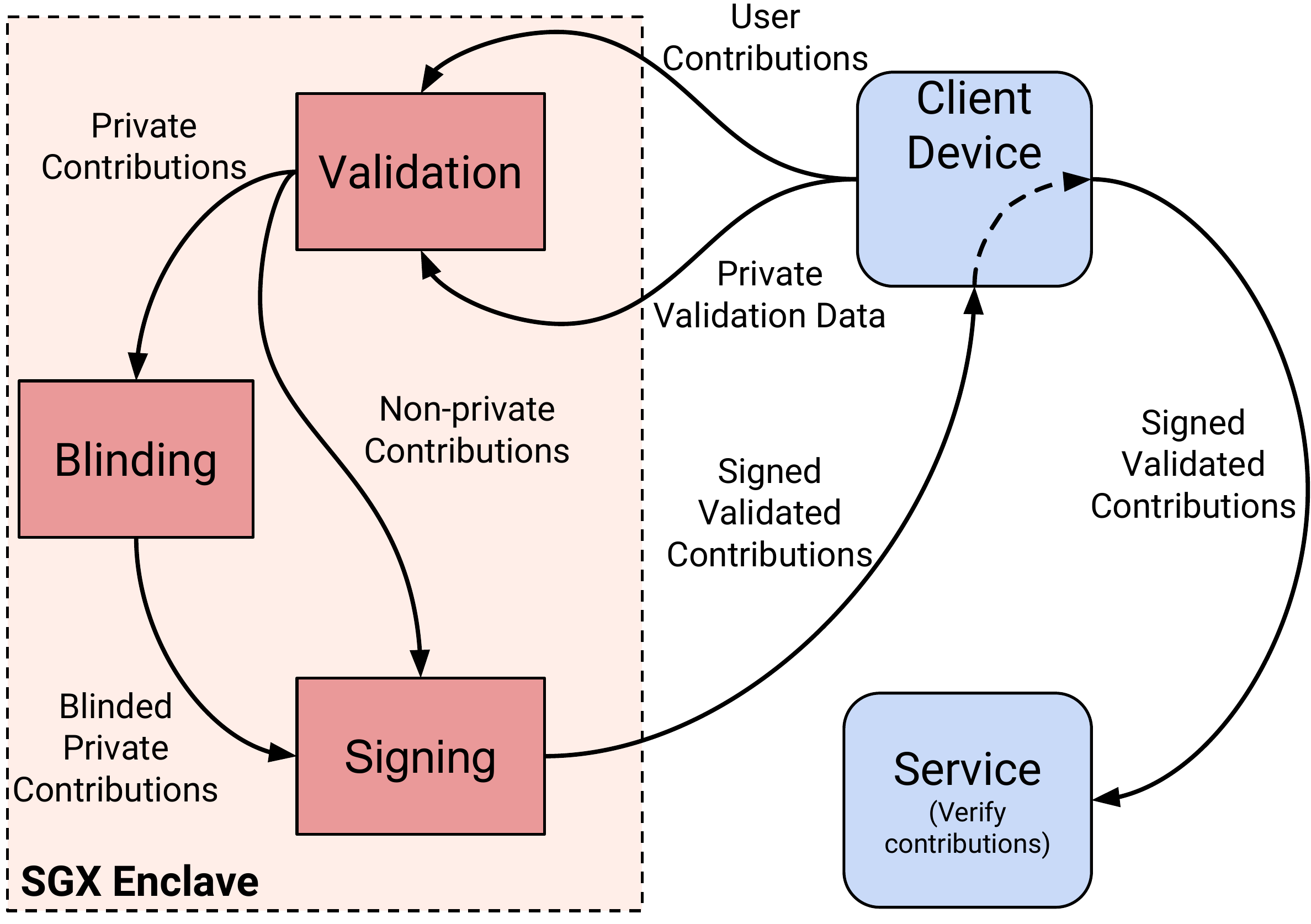}{SGX Glimmer design}{fig:glimmer-sgx}{1.0}{-20pt}

Figure~\ref{fig:glimmer-sgx} describes the design of an SGX-based Glimmer.  The
Glimmer has 3 main components.  A Validation component takes two types of input
from the client device: user contributions, which will be sent to the service
and private validation data, which is used internally by the Glimmer to run the
validation predicate.  In some cases, user contributions are used in the
validation predicate, while in other cases it will use the private data to validate the contributions.
For example, in the predictive keyboard service, model parameters are
range-checked against the valid weight range $[0,1]$.  In contrast, more
invasive validation predicates could request additional data, not contributed to
the service, such as individual key presses and timings or web browser logs
showing the typed data in user-triggered HTTP GET requests to corroborate the user contributions.  In the photos for maps, the Glimmer could request validation
information such as exact GPS location and tracks, a fingerprint of the camera
hardware, and access to other photos on the device to establish context.
Note that the Glimmer cannot directly obtain such information; it must
request this information from the host system.

Blinding is the second Glimmer component. Its purpose is to hide private,
user contributed values so that the service can compute aggregates on them
without revealing individual contributed data. To illustrate how this could work, we give a simple example~\cite{bonawitz2016practical}.  Assume the existence of a trusted blinding service -- which could,
itself, be implemented as a separate enclave on one of the clients, or as a
distinct trusted service -- that computes $N$ random blinding values $p_i$ such
that $\sum_{i=0}^{N-1} p_i = 0$. It then seals each $p_i$ value to the Glimmer code, and encrypts one of the sealed values to each of $N$ clients' public keys, distributing the encrypted blinding values to the
Glimmers running on each client. Each Glimmer for user $i$ can decrypt and unseal its blinding value. The Blinding component then computes the blinded user contribution
$y_i = x_i + p_i$. $y_i$ is safe to send to the service, since the service
cannot compute the private $x_i$ from it (because the blinding value $p_i$ is
secret). However, once the service sums all $y_i$'s together, it can compute the
accurate sum of all $x_i$'s: $\sum_{i = 0}^{N-1} y_i = \sum_{i=0}^{N-1} x_i +
\sum_{i=0}^{N-1} p_i = \sum_{i=0}^{N-1} x_i$. Recall that non-private
user contributions need not be blinded; for instance, in the crowd-sourced
photos for map locations, user-contributed photos are meant to be shared, so
they do not need to be blinded.

The third Glimmer component, Signing, takes a user-contributed input (blinded or
unblinded) and the result of the Validation component, which can be a boolean ``valid'' / ``invalid'', or a confidence value. If validation passed, the Signing component signs the
user-contributed input and returns it to the client for transmission to the
service. The signing key used can be provided by the service, and sealed (using
the SGX sealing facility) to the Glimmer code, so that it is only available to
instances of Glimmer enclaves. The signed contributions are passed back to the
client device, which can forward them to the service.

One last requirement is that the Glimmer convince both the user and service that it is correct -- i.e., that it has both input confidentiality and integrity properties. To convince the user, we envision vetting and formal verification by a third-party, such as the EFF; while the service could perform its own vetting and verification to convince itself.  Once it has been vetted, the hash of the Glimmer is published, and the user can use SGX to attest that their client is running the approved Glimmer.  Similarly the service can ensure that signing keys are sealed to the approved Glimmer.  Because the Glimmer is, necessarily,
small and limited in its external interactions, it is amenable to formal verification for absence of runtime errors such as buffer
and integer overflows~\cite{framac,uberspark}. Furthermore, much research has
been recently devoted to verifying formally the confidentiality of secret values
in SGX
enclaves~\cite{Sinha:2015:MVC:2810103.2813608,Sinha:2016:DVM:2908080.2908113}.
The burden on programmers is relatively low: programming in a simple programming
language (e.g., C) with relatively low-complexity idioms (e.g., bounded loops,
no function pointers, etc.), explicitly marking secret inputs, explicitly
marking declassification functions (e.g., blinding and encryption). Simple functional property verification can establish that every signed value has been validated and that no private information leaves the Glimmer without being blinded.

We have shown all components in Figure~\ref{fig:glimmer-sgx} within a single SGX
enclave, which is more efficient as there is only one transition in and out of
the enclave.  However, to increase ease of verification, the Glimmer can be decomposed so that each component runs in its own enclave.  Naturally, communication between components must now also be secured.

\section{Glimmer++}

\subsection{Validation confidentiality}

So far, we have described the use of Glimmers in applications where user contributions are used for a shared service.  However, Glimmers have applications to other problems where privacy and trust are at odds.  For example, consider the case of bot detection in a web service.  While CAPTCHAs are a standard method for detecting bots, they have their drawbacks, such as vulnerability to computer vision and CAPTCHA farms, and annoyance to legitimate human users.  An alternative solution is embedding a Javascript ``detector'' in the web page that heuristically detects whether a bot or a human is present.  Such solutions collect a large set of signals, such as how faithfully the client executes Javascript, fingerprints of the client's system software and hardware, and the timing and frequency UI interactions such as mouse movements and changes in focus~\cite{park_securing_2006,Gummadi:2009:NIS:1558977.1558998}.  These signals are sent back to the web service, which uses them to determine if the sender is a bot or a human.  However, these signals often contain private  information, such as the user's cookies, browsing history and browsing interests~\cite{JJLS10}. A Glimmer can protect individual privacy by performing the validation, which requires access to sensitive information locally on the client machine, and sending only 1 bit of information -- whether the user is human or not -- back to the web service.  

In such an adversarial example, the web service may wish to hide the exact validation predicate from the adversary, a property we call {\bf Validation Confidentiality}.  Glimmers can provide validation confidentiality by accepting encrypted code and data from the web service and decrypting and running that code inside the enclave where the plain text code is protected from observation by the hardware TEE.

One challenge is to make sure that the keys used to sign and encrypt the code and data are transfered securely to the Glimmer and that the Glimmer only accepts keys from a legitimate web service. This can be accomplished using remote attestation, which enables data, such as Diffie-Hellman (DH) handshake values, to be bound to code running in an enclave.  This would assert to the service that the DH handshake is occurring with a legitimate Glimmer.  Similarly, the Glimmer would need to ensure that the DH handshake is occurring with a legitimate service, which can be accomplished by the service signing its DH handshake values and embedding the signature verification key in the Glimmer code.  Once shared secrets are negotiated with DH key exchange, secret code and data can be securely transfered from the service to the Glimmer.

The other challenge is to prove input confidentiality to the user when part of the Glimmer can no longer be audited because it is encrypted and set dynamically at runtime.  This can be done by making the message format between the Glimmer and the service public, and having a runtime auditor check that each message is well formed and contains only one bit of information (i.e., a single bit plus a well-defined signature and challenge response).  While this does not preclude a covert channel, it puts a hard upper bound on the capacity of such a channel.

\subsection{Glimmer-as-a-service}

So far we have proposed that Glimmers run on client devices.  However, given the increasing trend towards Internet of things (IoT) devices, there are likely to be some devices that will make user contributions that must be trustworthy,  but do not have a processor with trusted computing capabilities.  In this case, we envision that a neutral third party may supply the capability to run a Glimmer.  This third party could simply be another device owned by the same user (such as a set-top box or home service), a local group of people that the user knows (such as their University, community or church), or even a well-known entity that is willing to sell or provide resources to improve user privacy (such as the EFF).  

The main criterion is that the client device needs to establish that it is sending its private data to a genuine Glimmer.  Fortunately, this can be accomplished using the same attestation mechanism to establish a secure channel as described above.  Attestation enables the client to be assured that the other endpoint of a secure communication channel is within an approved Glimmer.  Once this is done, the client can transmit the user contribution and private data and receive in return a signed (and if necessary, blinded) user contribution, which it can then forward to the service.

\section{Conclusion}

We proposed Glimmers of Trust, implemented on trusted computing hardware, that can provide trustworthiness guarantees of user-contributed data to services without compromising user privacy.  We describe a design using Intel SGX, which we are currently implementing.  While many previous proposals for SGX are for server-side uses~\cite{arnautov_scone:_2016,schuster_vc3:_2015,hunt_ryoan:_2016}, or confer mainly server-side benefits (i.e., DRM, mobile payments), we see Glimmers as one of the first uses of client-side trusted computing that can benefit {\em both} services and users.

\section*{Acknowledgements}

The ideas presented in this paper were inspired by a number of preliminary discussions with \'{U}lfar Erlingsson.

{\footnotesize \bibliographystyle{acm}
\bibliography{biblio}}

\end{document}